\documentclass[onecolumn]{IEEEtran}
\usepackage{amsmath}
\usepackage{theorem}
\usepackage{amsmath}
\usepackage{mathrsfs}
\usepackage{accents}
\usepackage{setspace}
\usepackage{url}
\usepackage{algorithm}
\usepackage{algorithmic}
\usepackage{epsfig}
\usepackage{amsfonts}
\usepackage{amssymb}
\usepackage{mathrsfs}
\usepackage{euscript}
\usepackage{graphicx}
\usepackage{multirow}
\usepackage{cite}
\newtheorem{theorem}{Theorem}

\usepackage[T1]{fontenc}
\usepackage[USenglish,UKenglish,french,spanish,italian]{babel}
\usepackage[nodayofweek,level]{datetime}
\newcommand{\mydate}{\formatdate{30}{11}{2016}}

\graphicspath{{./}{figures/}}

\ifCLASSINFOpdf
\else
\fi
\begin{document}

\begin{titlepage}

\begin{tabular}{l        r}

\includegraphics[bb=20bp 00bp 500bp 450bp,clip,scale=0.3]{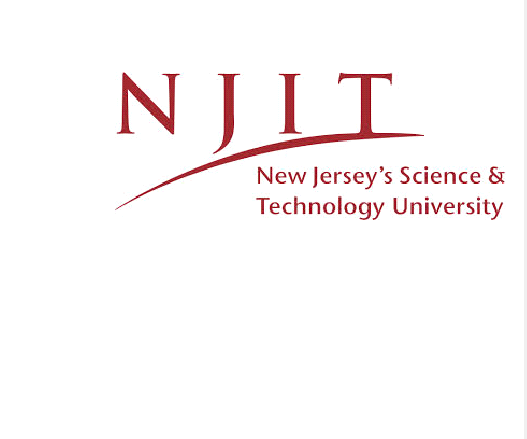} \hspace{6cm} & \includegraphics[bb=0bp -200bp 500bp 550bp,clip,scale=0.2]{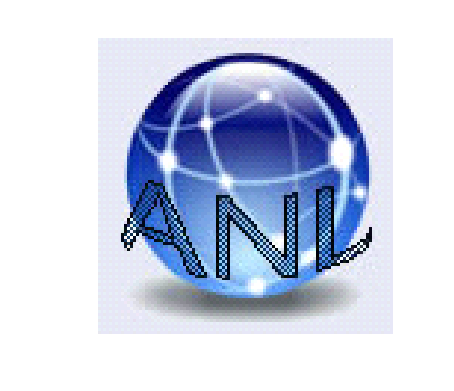}

\end{tabular}

\begin{center}

\textsc{\LARGE Towards Hierarchical Mobile Edge Computing: An Auction-Based Profit Maximization Approach}\\[1.5cm]

{\Large \textsc{Abbas Kiani}}\\ 
{\Large \textsc{Nirwan Ansari}}\\ 
[2cm]

{}
{\textsc{TR-ANL-2016-002}\\
\selectlanguage{USenglish}
\large \mydate} \\[3cm]

{\textsc{Advanced Networking Laboratory}}\\
{\textsc{Department of Electrical and Computer Engineering}}\\
{\textsc{New Jersy Institute of Technology}}\\[1.5cm]
\vfill

\end{center}

\end{titlepage}


\selectlanguage{USenglish}
\begin{spacing}{2}
\begin{abstract}
The multi-tiered concept of Internet of Things (IoT) devices, cloudlets and clouds is facilitating a user-centric IoT. However, in such three tier network, it is still desirable to investigate efficient strategies to offer the computing, storage and communications resources to the users. To this end, this paper proposes a new hierarchical model by introducing the concept of \textit{field}, \textit{shallow}, and \textit{deep} cloudlets where the cloudlet tier itself is designed in three hierarchical levels based on the principle of LTE-Advanced backhaul network. Accordingly, we explore a two time scale approach in which the computing resources are offered in an auction-based profit maximization manner and then the communications resources are allocated to satisfy the users' QoS.
\end{abstract}

\section{Introduction}\label{sec:Introduction}
The fog computing paradigm~\cite{bonomi2011connected} was introduced by Cisco as a new platform in which the goal is to support the requirements of Internet of Things (IoTs) varying from low latency, mobility, geo-distribution and location awareness\cite{bonomi2012fog}.
To this end, the fog computing platform was designed as a multi-tiered architecture in which different parts of an IoT application can be deployed on the IoT device, fog platform and a data center as three different tiers.
In the past few years, several efforts have developed similar concepts to the fog computing. Most notably, three years before the introduction of fog computing, the idea of cloudlet as a trusted, resource-rich computer which is well-connected to the Internet and available for use by nearby mobile devices was introduced in~\cite{satyanarayanan2009case}. The notion of the cloudlet or a "data center in a box" has been further developed by a research team at Carnegie Mellon University by introducing and developing various mechanisms~\cite{clinch2012close,satyanarayanan2013role,lewis2014cloudlet,satyanarayanan2015edge,ha2015adaptive}.

In parallel with the development of fog computing and the cloudlet concept, the so called Mobile Edge Computing (MEC) idea has being standardized by an Industry Specification Group (ISG) lunched by the European Telecommunications Standards Institute (ETSI)~\cite{hu2015mobile}. MEC  recognized as one of the key emerging technologies for 5G networks aims at providing computing capabilities in proximity of Mobile Users (MUs) and within the Radio Access Network (RAN), thereby reducing the latency and improving the Quality of Service (QoS)~\cite{hu2015mobile}. Moreover, MEC is becoming an important enabler of consumer-centric IoT with potential applications such as
smart mobility, smart cities, and location-based services~\cite{corcoran2016mobile,sun2}. Therefore, in such user-centric IoT concept in which the users participate in sensing and computing tasks, computation-intensive tasks still need to be offloaded to either the cloud or the computing resources at the edge.

In a MEC environment, a mobile subscriber/user can be considered as a person/entity with one or more IoT devices that can utilize the computing and storage capabilities at the edge. However, it is still desirable to investigate an efficient strategy that can be used to offer the computing and storage facilities, and accordingly the required communications bandwidth to a mobile subscriber. Such strategy not only has to
allow the users to adapt their computing and communications capacities according to their requirements but also has to change its economics by allowing the users to pay only for the resources that they utilize. In this regard, the main challenge is the resource poverty at the edge
where we are dealing with resource-poor computing facilities not big data centers. To this end, the current study aims to address the aforementioned issue by proposing an auction-based profit maximization approach.

\textbf{Contributions:} We have made three major contributions. 1) We propose a HIerachichal Mobile Edge Computing (HI-MEC) architecture in accordance with the principles of LTE-Advanced backhaul network and introduce the notion of \textit{field}, \textit{shallow} and \textit{deep} cloudlets.
2) We propose a two time scale mechanism to allocate the computing and communications resources to the MUs. The importance of the proposed two time scale is due to the fact that the economics of computing resources cannot change as quickly as the traffic loads of the MUs. In particular, the decision about the price and distribution of the computing resources are made in longer time frames, while the bandwidth allocations are updated in shorter time slots. To this end, we formulate a Binary Linear Programming (BLP) aimed at maximizing the profit of the service provider and a convex optimization problem for bandwidth allocation. We also design heuristic algorithms to solve the BLP problem and a centralized solution is proposed for the bandwidth allocation problem.
3) We evaluate the performance of the heuristic algorithms via extensive simulations.

\begin{figure*}\label{fig:1}
\includegraphics[width=17cm,height=9cm]{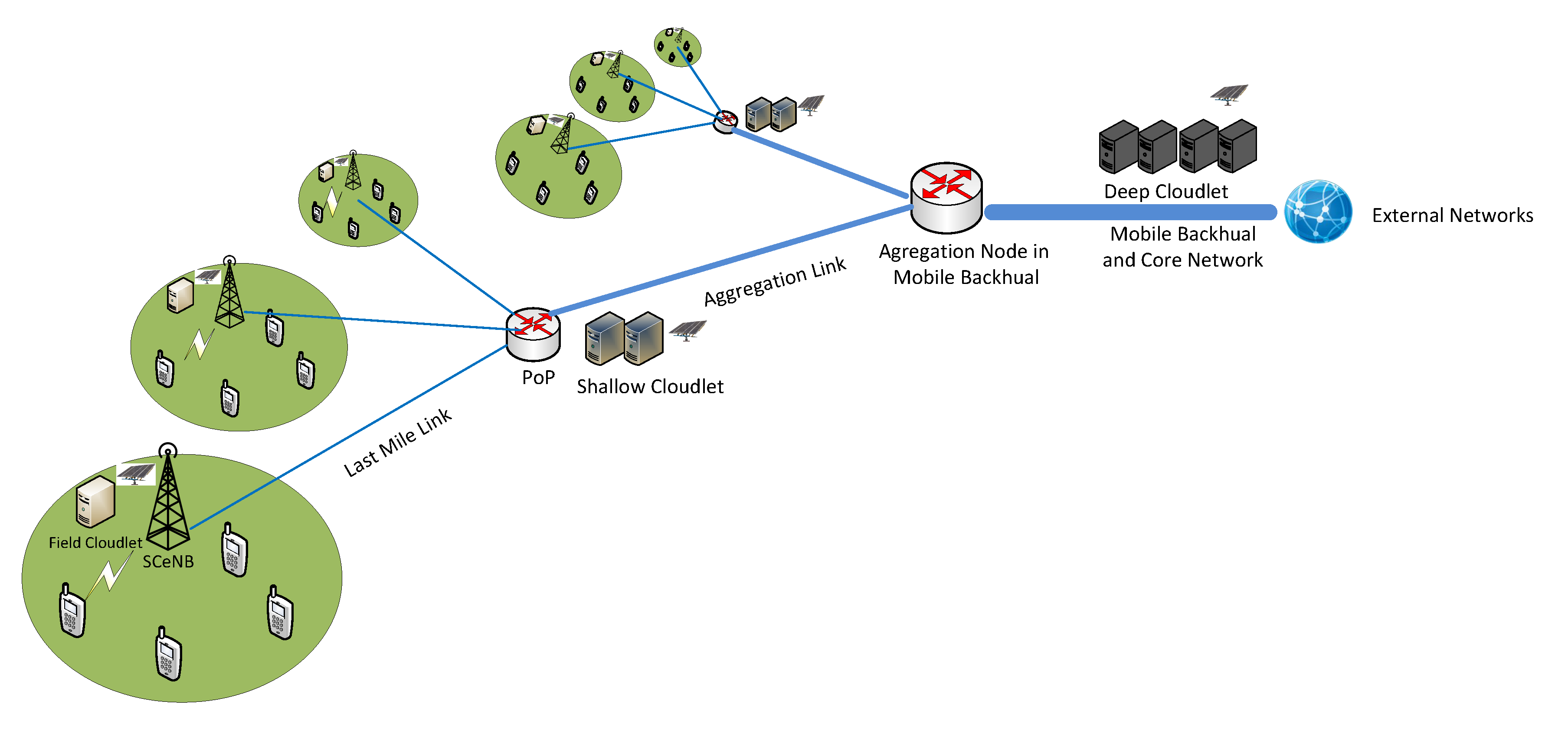}
\caption{System Model.}
\label{fig:1}
\end{figure*}
\textbf{Related Work:} In the past few years, a large and cohesive body of work investigated the major limitations of Mobile Cloud Computing (MCC), e.g., the radio access associated energy consumption of mobile devices and the latency experienced over Wide Area Network (WAN), and
the researchers came up with a variety of policies and algorithms. For instances, the computation offloading problem via joint optimization of the communication and computation resources is explored in~\cite{6923537} and a message-passing approach for the same problem is proposed in~\cite{khalili2016inter}. In~\cite{7511131}, a new cloudlet network architecture that brings the computing resources from the centralized cloud to the edge is proposed and the problem of Avatar, a software clone located in a cloudlet, migration to maintain a low E2E delay is investigated.
Recently, a cloudlet network planning approach for mobile access networks is introduced in~\cite{ceselli2015cloudlet} which optimally places the cloudlet facilities among a given set of available sites and then assigns a set of access points to the cloudlets by taking into consideration of the user mobility.

As discussed earlier, this paper proposes an auction-based profit maximization approach.
The idea is to apply the useful cloud-based strategies into the context of MEC by taking into consideration of the user mobility and the resource poverty at the edge. Specifically, our auction model is inspired by a model presented in~\cite{lampe2012maximizing} that proposes a concurrent Virtual Machine (VM) pricing and the distribution of VM instances across Physical Machines (PMs) in a data center.

The rest of the paper is organized as follows. Sections~\ref{Sec:Model} and~\ref{Sec:formulation}
describe the system model and problem formulation. We propose our auction-based profit maximization problem and the corresponding heuristic algorithms in~\ref{Sec:profit}. The bandwidth allocation problem and its centralized solution are presented in~\ref{Sec:bandwidth}. Finally, Sections~\ref{sec:simulations} and~\ref{sec:conclude} present numerical results and conclude the paper, respectively.

\begin{table*}
\centering
 \caption{Description of Symbols}\label{table1}
 \begin{tabular}{|p{2cm}|p{13cm}|}
  \hline
  \textbf{Symbols} & \multicolumn{1}{|l|}{\textbf{Description}}\\
  \hline\hline
  \multicolumn{2}{|c|}{\textbf{Provider Side}}\\
  \hline
  $\mathcal{A}\subseteq\mathbb{N}$  & Set of provisioned SCeNBs as the APs \\
  \hline
   $\mathcal{C}\subseteq\mathbb{N}$ & Set of all cloudlets \\
  \hline
  $\mathcal{C}_{f}\subseteq\mathcal{C}$ & Set of field cloudlets\\
  \hline
 $\mathcal{C}_{s}\subseteq\mathcal{C}$ & Set of shallow cloudlets \\
  \hline
  $c_{d}\in\mathcal{C}$ & Deep cloudlet. \\
  \hline
  $\mathcal{A}_{c_s}\subseteq\mathcal{A}$ & Set of APs connected to shallow cloudlet $c_s\in\mathcal{C}_s$ \\
  \hline
  $\mathcal{C}_a\subseteq\mathcal{C}$ & Set of cloudlet locations connected to AP $a\in\mathcal{A}$ \\
  \hline
   $\mathcal{V}\subseteq\mathbb{N}$ & Set of offered VMs  \\
  \hline
   $\mathcal{P}\subseteq\mathbb{N}$ & Set of available types of PMs \\
  \hline
  $\mathcal{P}_c\subseteq\mathcal{P}$ & Set of available types of PMs at cloudlet location $c\in\mathcal{C}$ \\
  \hline
   ${M}_{c}^{p}\subseteq\mathbb{N}$ & Available number of PMs of type $p\in\mathcal{P}$ at cloudlet $c\in\mathcal{C}$\\
  \hline
 $\mathcal{R}\subseteq\mathbb{N}$ & Set of resource types such as memory \\
  \hline
    $D^v$ & Maximum allowed data transfer to/from VM type $v\in\mathcal{V}$ within a time frame \\
  \hline
   $r_{min}^{v}$ & Base bandwidth of VM type $v\in\mathcal{V}$ \\
  \hline
 $RD_{r}^{v}$ & Resource demand of VM type $v\in\mathcal{V}$ for resource type $r\in\mathcal{R}$\\
  \hline
$RS_{r}^{p}$ & Resource supply of PM type $p\in\mathcal{P}$ for resource type $r\in\mathcal{R}$ \\
  \hline
$R_a$ & Capacity of the last mile link between AP $a\in\mathcal{A}$ and its connected shallow cloudlet\\
  \hline
 $R_{c_s}$ & Capacity of the aggregation link between shallow cloudlet $c_s\in\mathcal{C}_s$ and the aggregation node\\
  \hline
 $R_{c_d}$ & Capacity of the backhual link which connects the aggregation node to the deep cloudlet\\
  \hline
  \multicolumn{2}{|c|}{\textbf{Demand Side}}\\
  \hline
  $\mathcal{B}$ & Set of bids submitted for all types of VMs \\
  \hline
  $\mathcal{B}_c\subseteq\mathcal{B}$ & Set of $b\in\mathcal{B}$ that can be served at $c\in\mathcal{C}$\\
  \hline
 $\mathcal{B}^v\subseteq\mathcal{B}$ & Set of bids submitted for VM type $v\in\mathcal{V}$\\
  \hline
  $\mathcal{B}_a\subseteq\mathcal{B}$ & Set of bids submitted at AP location $a\in\mathcal{A}$\\
  \hline
  $\mathcal{B}_a^v\subseteq\mathcal{B}$ & Set of bids submitted for VM type $v\in\mathcal{V}$ at AP location $a\in\mathcal{A}$\\
  \hline
  $(1,...,|\mathcal{B}_a^v|)$ & Sequence of bids $b\in\mathcal{B}_a^v$ in a decreasing order of the corresponding prices\\
  \hline
$a_{b}$ & AP location of $b\in\mathcal{B}$ \\
  \hline
  $T_{b}$ & Desired VM type of $b\in\mathcal{B}$ \\
  \hline
  $k_{b}$ & Rank of $b\in\mathcal{B}$ in the corresponding sequence $(1,...,|\mathcal{B}_{a_b}^{T_b}|)$ \\
  \hline
 $e_{k,a}^v$ & Corresponding willingness price of the $k$th bid in $(1,...,|\mathcal{B}_{a}^{v}|)$\\
  \hline
 \multicolumn{2}{|c|}{\textbf{Profit}}\\
  \hline
  $x_{k,a}^{v}\in\{0,1\}$ & Binary decision variable that indicates whether the $k$th bid in sequence $(1,...,|\mathcal{B}_{a}^{v}|)$ is served or not. $x_{k,a}^{v}=1$ if the $k$th bid is served, and $x_{k,a}^{v}=0$ otherwise\\
  \hline
 $y_{m,c}^{p}\in\{0,1\}$ & Binary decision variable that indicates whether the $m$th PM of type $p\in\mathcal{P}$ at cloudlet $c\in\mathcal{C}$ is on or not. $y_{m,c}^{p}=1$ if the $m$th PM is on, and $y_{m,c}^{p}=0$ otherwise\\
  \hline
  $z_{b,m,c}^{p}\in\{0,1\}$ &  Binary decision variable that indicates the assignments of bid $b\in\mathcal{B}$ to the $m$th PM of type $p\in\mathcal{P}$ at cloudlet $c\in\mathcal{C}$. $z_{b,m,c}^{p}=1$ if bid $b\in\mathcal{B}$ is assigned to $m$th PM of type $p\in\mathcal{P}$ at cloudlet $c\in\mathcal{C}$, and $z_{b,m,c}^{p}=0$ otherwise\\
  \hline
    $q_c$ & Cost of electricity at cloudlet location $c\in\mathcal{C}$\\
  \hline
$P_{idle}^{p}$ & Idle power consumption of PM $p\in\mathcal{P}$ \\
  \hline
$P_{peak}^{v}$ & Average peak power consumption of a VM type $v\in\mathcal{V}$ \\
  \hline
$E_{usage}$ & Total power consumption (including that of network facilities) divided by the power consumption at the cloudlets \\
  \hline
\end{tabular}
\end{table*}

\section{System Model}\label{Sec:Model}
Fig.~\ref{fig:1} shows our proposed HI-MEC architecture designed for provisioning mobile edge computing services by an edge-computing service provider (a service provider in short). Based on the principles of LTE-Advanced backhaul network~\cite{alliance2012small}, we introduce the notion of \textit{field}, \textit{shallow} and \textit{deep} cloudlets. In particular, in a HI-MEC environment, we have several field cloudlets as the resource-poor facilities co-located with Small Cell enhanced Node Bs (SCeNBs). The shallow cloudlets as the resource-middle class facilities are also hosted at the first level of aggregation nodes, i.e., at Point of Presences (PoPs). Moreover, in order to leverage the resource-rich facilities, we consider one deep cloudlet for each HI-MEC enviroment located at mobile backhual. In the proposed hierarchical model, each SCeNB is assumed to be connected to one PoP using a dedicated last mile link. Moreover, there is a dedicated aggregation link between each PoP and the aggregation node. In other words, each field cloudlet has access to only one shallow cloudlet connected via a dedicated last mile link, and all shallow cloudlets are connected to the deep cloudlet via aggregation links and mobile backhual.

We assume that the network has been optimally designed in terms of the connections of the SCeNBs to the PoPs by taking into consideration of different parameters like link lengths and capacities. A list of the most symbols is
summarized in Table~\ref{table1}. However, in order to ease the reading, the symbols used in Sections~\ref{subsec:heuristic} and~\ref{Sec:bandwidth} are not included in this table and are explained in the corresponding sections.

We consider a two time scale model in which the running time of the HI-MEC environment is divided into a
sequence of time frames at equal length, $T$, e.g., five minutes. Each time frame itself is also divided into a sequence of time slots at equal length, $\tau$, e.g., a few seconds.
Our goal is to maximize the service provider total profit during the time
frame $T$ and minimize the total delay experienced by the users during the time slot $\tau$. Note that for the analysis, we
consider a single time frame, e.g., $\Delta$ as the time frame of interest (or a single time slot, e.g., $\delta$ as the time slot of interest) and omit the explicit time dependence in the notations through the paper.

\subsection{Provider Side}
The service provider provides the MUs (users in short) by a set of computing and communications facilities as an augmentation to their mobile device capacities. The computing facilities are provisioned as different types of Virtual Machines (VMs) running on Physical Mashines (PMs) located at different cloudlet sites. To manage the fluctuations in user demands while taking into consideration of the limitations of available resources at the edge, the service provider should consider a flexible pricing methods in which the resources are priced according to the demands. To this end, we consider an auction-based pricing model such as Amazon's Elastic Compute Cloud (EC2) spot pricing~\cite{EC2,lampe2012maximizing,zheng2015bid}. In such strategy, the service provider updates the prices for each type of VM at the beginning of each time frame that depend on the available resources and demands. The minimum granularity in offering the computing resource is assumed to be one VM instance in one time frame.
The service provider also renders the required communications bandwidth between the users and the VMs, i.e., the SCeNBs as the Access Points (APs) as well as the network connection between the APs and the cloudlet locations.

\subsection{Demand Side}
\vspace{-.09cm}
The service provider tenders the communications and edge-computing facilities as a service to the MUs. The MUs can benefit from the provided service, e.g., by offloading their mobile applications, and hereby prolong their device battery life-time. However, the users must submit their demand bids for the offered service stating their maximum willingness price for their desired VM type. The maximum willingness price can be decided using the spot price history. We assume that the users can submit their bids at any time but the service provider runs the auction at the beginning of each time frame in which the bids above the spot price are served, and those below the spot price are rejected.
In fact, it is assumed that the demand bids are submitted based on the required VM type but the service provider will guarantee communications bandwidth for the served bids.
Without loss of generality, if a user demands more than one instance of a specific type of VM type, we treat the requested instances as different bids but with the same maximum willingness price.

\section{Problem Formulation}\label{Sec:formulation}
The service provider not only has to decide the final price, which depends on the number of served bids for each type of VM, but also has to determine the assignments of the VMs among the cloudlet locations such that the communications requirements are also guaranteed. To this end, we propose an auction-based profit maximization problem to be solved by the service provider.
The profit gained by running the proposed HI-MEC environment is assumed to be given by the revenue of serving the VM demands minus the electricity cost of running the computing and network facilities, and the revenue lost due to network delay.

\subsection{Revenue}
The revenue of the service provider in a time frame depends on its decision about the spot price for each type of VM. We consider a local pricing approach in which the price for a specific type of VM varies from one AP location to another AP depending on the demand and supply but all the served bids in one AP location pay an identical price, i.e., equilibrium price per instance of a VM type.  On the other hand, at each AP location, for a given type of VM, only those bids whose respective prices are greater than or equal to the equilibrium price can be served with their desired VM instances. We thus establish the revenue of the service provider in one time frame as,

\begin{equation}\label{equ1}
R=\sum_{a\in\mathcal{A}}\sum_{v\in\mathcal{V}}\sum_{k=1}^{|\mathcal{B}_a^v|}x_{k,a}^v(k*e_{k,a}^v-(k-1)*e_{k-1,a}^v)
\end{equation}
where we assume that the binary variables $x_{k,a}^v$ are decided such that $x_{k,a}^v\leq x_{k-1,a}^v$. In the presented definition for revenue, for example, at AP location $a$, the final local price for one instance of VM type $v$, is set to the maximum willingness price of the last served bid in sequence $(1,...,|\mathcal{B}_{a}^{v}|)$. In other words, all the bids with willingness prices above this bid are served, and on the other hand, all the bids with willingness prices below this bid are rejected. The total revenue is thus calculated by summing over all the bids in sequence $(1,...,|\mathcal{B}_{a}^{v}|)$ with consideration of their willingness prices ($e_{k,a}^v$). Going from the $(k-1)$th bid to the $k$th bid, if ($x_{k,a}^v=1$), the new revenue, $k*e_{k,a}^v$, is added to the summation and the previous revenue, $(k-1)*e_{k-1,a}^v$, is deducted from the summation.

\subsection{Electricity Cost}
The electricity cost of the service provider depends on different variables like  the number of turned on PMs at each cloudlet and the distribution of the VMs among the PMs.
Following the power consumption model adapted for data centers~\cite{PUE,ghamkhari2013energy,kiani2015profit,kiani2016fundamental}, the total electricity cost (EC) in one time frame can be computed as,
\begin{equation}\label{equ2}
\begin{split}
EC=TE_{usage}(\sum_{b\in\mathcal{B}}\sum_{c\in\mathcal{C}_{a_b}}\sum_{p\in\mathcal{P}_c}\sum_{m=1}^{min(|\mathcal{B}|,M_{c}^{p})}q_cz_{b,m,c}^{p}P_{peak}^{T_b}\\ +\sum_{c\in\mathcal{C}}\sum_{p\in\mathcal{P}_c}\sum_{m=1}^{min(|\mathcal{B}|,M_{c}^{p})}q_c y_{m,c}^{p}P_{idle}^{p})
\end{split}
\end{equation}
where the first term corresponds to the electricity cost of VMs' power consumption and the second term is to consider the related cost of PMs' idle power consumption.

\subsection{Lost Revenue}
The proposed architecture is a MEC architecture where the users expect to experience a low latency connecting to their VMs. Therefore, for QoS satisfaction, we incorporate a lost revenue into our profit maximization problem due to the network delay experienced by the users.
The idea is to first serve the bids as close as possible to the edge, and then allocate bandwidth to those bids that have to be served at a shallow/deep cloudlet due to high demands at their corresponding AP locations. In other words, field cloudlets have to be the first priority to serve a bid while shallow and deep cloudlet facilities have the second and third priorities, respectively.

Let $r_b$ be the bandwidth allocated to bid $b$ on all the links that it has to go through. For example, if bid $b$ is served at the deep cloudlet, $r_b$ is allocated to bid $b$ on all corresponding last mile, aggregation and mobile bakchual links. In other words, there is a dedicated link of capacity $r_b$ between the corresponding AP of bid $b$ and its assigned cloudlet location.
Since the users are interested in their QoS, rather than their allocated bandwidth, we translate the allocated bandwidth to our lose revenue.

In a nutshell, at any time $t\in T$, we denote the traffic load of a given bid $b$ on its dedicated link, i.e., $r_b$, by $A_b(t)$. Therefore, within interval $T$ , bid $b$ makes its dedicated link busy for
$\frac{\int_{0}^{T} A_b(t)dt}{r_b}$ seconds. Thus, the link utilization for bid $b$ is $\frac{\int_{0}^{T} A_b(t)dt}{Tr_b}$.
Here, the network delay is related to the link utilization such that the less time is the link busy, the less network delay is experienced.
The total traffic load of a bid within a time slot is upper bounded by its maximum data transfer to/from the VM, i.e., $\int_{0}^{T} A_b(t)dt\leq D^{T_b}$. Moreover, we assume that the allocated bandwidth of each bid is lower bounded by the base bandwidth of its VM type, i.e., $r_b\geq r_{min}^{T_b}$.
Therefore, the link utilization of a bid is upper bounded with its maximum data transfer as well as the base bandwidth as follows,
\begin{eqnarray}\label{equ3}
\frac{\int_{0}^{T} A_b(t)dt}{Tr_b}\leq \frac{D^{T_b}}{Tr_{min}^{T_b}}
\end{eqnarray}
The idea is to incorporate this upper bound into our profit maximization which is solved every time frame and then update the bandwidth allocated to the bids every time slot based on the traffic loads. We thus define our lost revenue as,
\begin{eqnarray}\label{equ4}
LR=\sum_{a\in\mathcal{A}}\sum_{b\in\mathcal{B}_a}\sum_{c\in\mathcal{C}_a\setminus\mathcal{C}_f}\sum_{p\in\mathcal{P}}
\sum_{m\in M_{c}^{p}}\xi_{a,c}\frac{z_{b,m,c}^{p}D^{T_b}}{Tr_{min}^{T_b}}
\end{eqnarray}
where $\xi_{a,c}$ are the coefficients set by the service provider based on the importance of QoS compared to the profit and by taking into consideration of the link lengths between APs and their connected cloudlets.
Moreover, the reason behind using the upper bound is to derive a QoS satisfaction which is VM type oriented.
%

\section{Profit Maximization}\label{Sec:profit}
Note that users can submit or cancel their bids or change their willingness prices. Moreover, the AP location of a user changes when she moves to other location. Therefore, the service provider must update its decision on serving the bids periodically. To this end, we propose to maximize the auction-based profit at the beginning of each time frame.
\subsection{Binary Linear Programming}
The proposed optimization problem is formulated as,
\begin{eqnarray}
\underset{x_{k,a}^{v},~y_{m,c}^{p},~z_{b,m,c}^{p}}{\text{maximize}} (R-EC-LR)\nonumber
\end{eqnarray}
\begin{eqnarray}
\text{C1}:\sum_{c\in\mathcal{C}_{a_b}}\sum_{p\in\mathcal{P}_c} \sum_{m=1}^{min(|\mathcal{B}|,M_{c}^{p})}z_{b,m,c}^{p}=x_{k_b,a_b}^{T_b}
~\forall b\in\mathcal{B}\nonumber
\end{eqnarray}
\begin{eqnarray}
\text{C2}: \sum_{b\in\mathcal{B}}z_{b,m,c}^{p}RD_{r}^{T_b}\leq RS_{r}^{p}y_{m,c}^{p}~\forall p,m,c,r\nonumber
\end{eqnarray}
\begin{eqnarray}
\text{C3}: \sum_{b\in\mathcal{B}_a}\sum_{c\in\mathcal{C}_a\setminus\mathcal{C}_f}\sum_{p\in\mathcal{P}_{c}}\!
\sum_{m=1}^{min(|\mathcal{B}_a|,M_{c}^{p})} z_{b,m,c}^{p}r_{min}^{T_b}\leq R_a~\forall a\nonumber
\end{eqnarray}
\begin{eqnarray}
\text{C4}: \sum_{a\in\mathcal{A}_{c_s}}\sum_{b\in\mathcal{B}_a}\sum_{p\in\mathcal{P}_{c_d}}
\sum_{m=1}^{min(|\mathcal{B}_a|,M_{c_d}^{p})} z_{b,m,c_d}^{p}r_{min}^{T_b}\leq R_{c_s}~\forall c_s\nonumber
\end{eqnarray}
\begin{eqnarray}
\text{C5}: \sum_{b\in\mathcal{B}}\sum_{p\in\mathcal{P}_{c_d}}
\sum_{m=1}^{min(|\mathcal{B}|,M_{c_d}^{p})}z_{b,m,c_d}r_{min}^{T_b}\leq~R_{c_d}\nonumber
\end{eqnarray}
\begin{eqnarray}
\text{C6}: x_{k,a}^v\leq x_{k-1,a}^v~\forall v, a, 2\leq k \leq |\mathcal{B}_a^v|\nonumber
\end{eqnarray}
\begin{eqnarray}
\text{C7}: y_{m,c}^{p}\leq y_{m-1,c}^{p}~\forall c, p , 2\leq m\leq M_c^p\nonumber
\end{eqnarray}
\begin{eqnarray}
\text{C8}: x_{k,a}^{v}\in\{0,1\}~\forall v, a, 1\leq k\leq |\mathcal{B}_a^v|\nonumber
\end{eqnarray}
\begin{eqnarray}
\text{C9}: z_{b,m,c}^{p}\in\{0,1\}~\forall b,m,c,p\nonumber
\end{eqnarray}
\begin{eqnarray}\label{equ6}
\text{C10}: y_{m,c}^{p}\in\{0,1\}~\forall m,c,p\nonumber\\
\end{eqnarray}
where the objective is to maximize the profit defined as the $Revenue-Electricity Cost-Lost Revenue$. The equality constraint C1 in~(\ref{equ6}) is to ensure that the served bids are assigned to a PM at a cloudlet location connected to their AP locations. Inequality constraint C2 is also to lower bound the total resource demands of all the bids assigned to a PM by the resource supply of that machine. In addition, we use inequality constraints C3, C4 and C5 to bound the total minimum bandwidth of the bids traversing a link by the bandwidth capacity of that link. Note that C3, C4 and C5 are formulated for the last mile, aggregation and backhual links, respectively. Moreover, by inequality constraints C6, we enforce the requirement of our defined revenue function. Constraint C7 is designed to give priority to the PMs with lower running index at one cloudlet location over those with higher index at the same location. Finally, constraints C8, C9 and C10 are to restrict our variables to the binary choices.

\subsection{Heuristics}\label{subsec:heuristic}
While the proposed BLP optimization model offers flexibility, finding an optimal solution presents computational complexity. The complexity grows fast with the number of bids and PMs. In order to obtain high quality solutions in a reasonable time, we propose two heuristic algorithms that employ VM pricing and VM distribution techniques~\cite{lampe2012maximizing}. The pseudo codes for VM pricing and VM distribution algorithms are shown in Algorithms~\ref{alg:1} and~\ref{alg:2}, respectively. In fact, we follow a two phases approach.

In the first phase (Algorithm \ref{alg:1}), for each type of VM at each AP location, we first estimate the serving cost of one VM instance, i.e., $\varphi_a^v$, by taking a weighted average over all suitable type of PMs across all connected field, shallow, and deep cloudlets to that location. In our cost estimation, we consider both electricity cost and the lost revenue (lines 2-19). We then identify the favorable number of the bids to be served, i.e., $\hat{k}_a^v$ and the final local price, i.e., $\omega_a^v$ such that the estimated profit is maximized (lines 21-26). Finally, for each AP $a$ and VM type $v$, we store all those bids with a rank less than or equal to $\hat{k}_a^v$ in the set of served bids, i.e., $S$ (line 27).

In the VM distribution phase (Algorithm~\ref{alg:2}), we first initialize an instance count $m_p^c$ for each type of PM at each AP location (lines 1-5). We then
search the set of all the available PMs and the cloudlet locations to find a favorite PM, i.e., $\hat{p}$, at a favorite cloudlet, i.e., $\hat{c}$. For a given instance of a PM type at a given cloudlet, we scan the set of all the served bids and create a packing list for that machine, i.e., $L_c^p$. The packing list for a PM is created based on its resource constraints and the possibility of serving a bid at that PM. We subsequently compute the utility function for each PM at each cloudlet location, i.e., $u_c^p$. Accordingly, both the favorite PM type and cloudlet location are identified by comparing all the utility functions (lines 8-25), and all the bids in the corresponding packing list, i.e,  $L_{\hat{c}}^{\hat{p}}$, are assigned to one instance of $\hat{p}$ at $\hat{c}$ (lines 26-32). Finally, the assigned bids are removed from the set of served bids and this process is repeated until all the served bids are assigned or no suitable PM and cloudlet location is found for the VM assignment (lines 33-34). The complexity of the VM distribution presented in~Algorithm~\ref{alg:2} corresponds to $\mathcal{O}(|\mathcal{B}|^2\ast|\mathcal{P}|\ast|\mathcal{C}|)$.

\begin{algorithm}
\caption{VM pricing}
\label{alg:1}
\begin{algorithmic}[1]
\STATE $S\gets \emptyset$
\FORALL{$v\in\mathcal{V}$}
\FORALL{$a\in\mathcal{A}$}
\STATE $g_a^v\gets0, EC_a^v\gets0$
\FORALL{$c\in\mathcal{C}_a$}
\FORALL{$p\in\mathcal{P}_c$}
\IF{$p.canHost(v)=true$}
\STATE{$g^{p}_{a,c}\gets min(M_c^p, |\mathcal{B}_a^v|)$}
\IF{$c\in C_a\setminus C_f$}
\STATE{$g^{p}_{a,c}\gets min(g^{p}_{a,c}, \frac{R_a}{r_{min}^v})$}
\ENDIF
\STATE{$g_a^v\gets g_a^v+g^{p}_{a,c}$}
\STATE{$f^{p}_{a,c}\gets Tq_c(p_{peak}^v+\frac{p_{idle}^p}{|\mathcal{R}|}\sum_{r\in\mathcal{R}}\frac{RD_r^v}{RS_r^p})+\xi_{a,c}\frac{D^v}{r_{min}^v}$}
\STATE{$\varphi_a^v=\varphi_a^v+g^{p}_{a,c}*f^{p}_{a,c}$}
\ENDIF
\ENDFOR
\ENDFOR
\IF{$g_a^v>0$}
\STATE $\varphi_a^v\gets \frac{\varphi_a^v}{g_a^v}$
\STATE $\hat{\rho}_a^v\gets 0, \hat{k}_a^v=0, \omega_a^v\gets0$
\FOR{$k=1\rightarrow|B_a^v|$}
\STATE{$\rho_a^v\gets k*(e_{k,a}^v-\varphi_a^v)$}
\IF{$\rho_a^v\geq\hat{\rho}_a^v$}
\STATE{$\hat{\rho}_a^v\gets \rho_a^v,\omega_a^v\gets e_{k,a}^v, \hat{k}_a^v=k$}
\ENDIF
\ENDFOR
\STATE $S\gets S\cup\{b\in\mathcal{B}_a^v\mid k_b<=\hat{k}_a^v\}$
\ENDIF
\ENDFOR
\ENDFOR
\end{algorithmic}
\end{algorithm}

\begin{algorithm}
\caption{VM distribution}
\label{alg:2}
\begin{algorithmic}[1]
\FORALL{$c\in\mathcal{C}$}
\FORALL{$p\in\mathcal{P}_c$}
\STATE $m_c^p\gets 0$
\ENDFOR
\ENDFOR
\REPEAT
\STATE $\hat{p}\gets\emptyset, \hat{c}\gets\emptyset, \hat{u}\gets0$
\FORALL {$p\in\mathcal{P}$}
\FORALL{$c\in\mathcal{C}$}
\IF{$p\in\mathcal{P}_c$}
\IF{$m_c^p<M_c^p$}
\STATE $L_c^p\gets\emptyset$
\FORALL{$b\in S\cap \mathcal{B}_{c}$}
\IF{$c.canHost(a_b) \vee m_c^{p}.canHost(L_c^p\cup b)=true$}
\STATE $L_c^p\gets L_c^p\cup b$
\ENDIF
\ENDFOR
\STATE $u_c^p\gets\frac{\sum_{b\in L_c^p}\omega^{T_b}_{a_b}}{q_c(p_{idle}^p+\sum_{b\in L_c^p}p_{peak}^{T_b})+\sum_{b\in L_c^p}\frac{\xi_{a_b,c}D^{T_b}}{Tr_{min}^{T_b}}}$
\IF{$u_c^p>\hat{u}$}
\STATE $\hat{u}\gets u_c^p, \hat{p}\gets p, \hat{c}\gets c$
\ENDIF
\ENDIF
\ENDIF
\ENDFOR
\ENDFOR
\IF{$\hat{p}\neq\emptyset$}
\STATE $y_{m_{\hat{p}}^{\hat{c}},\hat{c}}^{\hat{p}}\gets1$
\STATE $m_{\hat{c}}^{\hat{p}}\gets m_{\hat{c}}^{\hat{p}}+1$
\FORALL{$b\in L_{\hat{c}}^{\hat{p}}$}
\STATE $z_{b,m_{\hat{c}}^{\hat{p}},\hat{c}}^{\hat{p}}\gets 1$
\STATE \textit{update the capacity of all links between $a_b$ and $\hat{c}$ according to $r_{min}^{T_b}$}
\ENDFOR
\ENDIF
\STATE $S\gets S\setminus L_{\hat{c}}^{\hat{p}}$
\UNTIL{$S=0 \vee \hat{p}=\emptyset$}
\end{algorithmic}
\end{algorithm}

\section{Bandwidth Allocation}\label{Sec:bandwidth}
Based on the VM assignment in the previous section, we now define an optimization problem to find the optimal bandwidth allocated to each served bid, i.e., $r_b$. Our goal is to minimize the total network delay experienced by the served users on the link between their corresponding APs and cloudlets. Note that the delay between a user and AP which is related to the radio resource allocation is not the focus of this paper since it has already been addressed in other studies such as~\cite{6923537}.
Let $\{1,...,N\}$ be the set of all bids served at a shallow or deep cloudlet. $a_b$ and $c_b$ are also the corresponding AP and cloudlet locations of bid $b$, respectively. Moreover, we define $\{1,...,M\}$ as the set of all the links in our HI-MEC environment including all the last mile, aggregation links and the mobile backhual link. Let $v_{mb}$ be a binary variable such that $v_{mb}=1$ if the traffic load of bid $b$ traverses link $m$.

\subsection{Convex Optimization}
We propose to solve the bandwidth allocation problem shown in~(\ref{equ7}) at the beginning of the time slot of interest. The objective of this optimization problem is to minimize the total delay experienced by the users who have been served at shallow cloudlets or the deep cloudlet location, by taking into consideration of the traffic load of each user at the beginning of the time slot of interest, i.e., $\lambda_b$. Constraints C1 and C2 are to bound the bandwidth allocated to bid $b$ by the the lower and upper boundary values $l_b$ and $u_b$, respectively. Note that these values are positive and decided by the service provider for example based on the VM types and the traffic loads. The lower bound $l_b$ is also upper bounded by the base bandwidth considered during the auction, i.e., $l_b\leq r_{min}^{T_b}$. Moreover, constraint C3 is to bound the bandwidth allocation by the physical bandwidth capacity of the links. In fact, the total bandwidth allocated to the bids traversing link $m$ is upper bounded by its capacity, i.e., $R_m$.
\begin{eqnarray}
\underset{r_b}{\text{minimize}} \sum_{b=1}^{N}\xi_{a_b,c_b}\frac{\lambda_b}{r_b}\nonumber
\end{eqnarray}
\begin{eqnarray}
\text{C1}: r_b\geq l_b~\forall b\in{1,...,N}\nonumber
\end{eqnarray}
\begin{eqnarray}
\text{C2}: r_b\leq u_b~\forall b\in{1,...,N}\nonumber
\end{eqnarray}
\begin{eqnarray}\label{equ7}
\text{C3}: \sum_{b=1}^{N}v_{mb} r_b\leq R_m~\forall m\in{1,...,M}\nonumber\\
\end{eqnarray}

\subsection{Centralized Optimal Solution}
The proposed bandwidth allocation problem is a convex optimization with $2N+M$ constraints. The complexity of this problem may increase as the numbers of the served bids and the links increase. However, a HI-MEC network is assumed to be limited by the number of the links and the computing capacity to serve as few as several thousand  bids. Therefore, it is desirable to derive a centralized optimal solution for this problem. To this end, we define the matrix $V=(v_{mb})_{M\times N}$ to show the traverse of the bids on each link based on our already defined binary variable $v_{mb}$. Let $R=(R_1,...,R_M)$ and $r=(r_1,...,r_N)$ also be the vectors of the capacity of the links and the bandwidth allocated to the bids, respectively. To derive the optimal solution, we apply the method of Lagrange multipliers since the constraints of Problem~(\ref{equ7}) are linear, and the Kuhn-Tucher conditions are necessary and sufficient for an existing optimal solution~\cite{boyd2004convex,guo2013cooperative}.

\begin{theorem}\label{theorem1}
There exists $\gamma_m\geq0$ ($m\in{1,...,M})$ such that $\forall b\in{1,...,N}$:
\begin{eqnarray}\label{equ8}
\hat{r_b}=\sqrt{\frac{\xi_{a_b,c_b}\lambda_b}{\sum_{m=1}^{M}\gamma_mv_{mb}}}\\
l_b\leq r_b \leq u_b\nonumber
\end{eqnarray}
and $\forall m\in{1,...,M}$
\begin{eqnarray}\label{equ9}
\gamma_m((V.r)_m-R_m)=0
\end{eqnarray}
where $\hat{r_b}$ is the optimal solution for Problem~(\ref{equ7}).
\end{theorem}
\begin{IEEEproof}
Our proof is based on the assumption that the bandwidth allocation space of Problem~(\ref{equ7}) is a nonempty, convex and compact set and thus  our objective function is strictly convex with respect to $r_b$. Then, we define $\alpha_b\geq0$ and $\beta_b\geq0$ $\forall b\in{1,...,N}$ as well as $\gamma_m\geq0$ $\forall m\in{1,...,M}$ as the Lagrange multipliers for constraints C1, C2 and C3 in problem~(\ref{equ7}), respectively. Therefore, the Lagrangian becomes,
\begin{eqnarray}\label{equ10}
\mathcal{L}(r,\alpha,\beta,\gamma)=\sum_{b=1}^{N}\xi_{a_b,c_b}\frac{\lambda_b}{r_b}+\sum_{b=1}^{N}\alpha_b(l_b-r_b)\nonumber\\
+\sum_{b=1}^{N}\beta_b(r_b-u_b)
+\sum_{m=1}^{M}\gamma_m((V.r)_m-(R)_m).
\end{eqnarray}
To optimize the objective by applying the necessary and sufficient conditions, we have
\begin{eqnarray}\label{equ11}
\Delta\mathcal{L}(\hat{r},\alpha,\beta,\gamma)=0\Leftrightarrow\nonumber
\end{eqnarray}
\begin{eqnarray}\label{equ12}
-\xi_{a_b,c_b}\frac{\lambda_b}{{r_b}^2}-\alpha_b+\beta_b+
\sum_{m=1}^{M}\gamma_mv_{mb}=0~\forall b\in{1,...,N}
\end{eqnarray}
and
\begin{eqnarray}\label{equ13}
\alpha_b(l_b-\hat{r_b})=0~\forall b\in{1,...,N},\nonumber\\
\beta_b(\hat{r_b}-u_b)=0~\forall b\in{1,...,N},\nonumber\\
\gamma_m((V.r)_m-(R)_m)=0~\forall m\in{1,...,M},
\end{eqnarray}
where $\hat{r}=(\hat{r_1},...,\hat{r_N})$ is the optimal solution to Problem~(\ref{equ7}). Noting the values of the Lagrange multipliers in~(\ref{equ13}) and focusing on the general case when $l_b<r_b<u_b$, one can conclude $\alpha_b=0$ and $\beta_b=0$. In fact, we are not interested in special cases when $r_b$ is equal to the boundary values. Therefore, by solving~(\ref{equ12}) for $\alpha_b=0$ and $\beta_b=0$, $\hat{r_b}$ is derived and the proof is complete.
\end{IEEEproof}
The result of Theorem~\ref{theorem1} indicates that the optimal bandwidth for each bid can be achieved by the optimal multipliers of its associated links.
For example, when a bid is served at the deep cloudlet, its optimal bandwidth can be solved by the optimal multipliers of its associated last mile and aggregation links as well as the mobile backhual link. Therefore, solving this problem in a distributed manner for the case that the numbers of bids and the links scale up can be investigated in a future work.

\begin{table*}
\center
\caption{Computation times comparison between heuristic and optimal approaches.}
\label{table2}
\begin{tabular}{|c|c|c|c|c|}
  \hline
    & 50 (bids) & 100 (bids) & 1000 (bids) & 2000 (bids)\\\hline
  Heuristic case 1 & 0.052 (s) & 0.79 (s) & 1.89 (s) & 5.55 (s) \\\hline
  Optimal case 1 & 2.17 (s)& 76.32 (s)& 107.65 (s)& 458.86 (s) \\\hline
  Heuristic case 2 & 0.31 (s)& 0.94 (s)& 2.716(s) & 5.75 (s) \\\hline
  Optimal case 2 & 2.53 (s)& 31.99 (s)& 97.34 (s)& 570.53 (s)\\
  \hline
  \end{tabular}
\end{table*}

\section{Simulation Results}\label{sec:simulations}
In this section, we compare the results of the heuristic VM pricing and VM distribution algorithms with the optimal results in solving the proposed profit maximization problem (BLP). We consider a HI-MEC environment consisting of five AP locations, each co-located with a field cloudlet, and ,two PoPs, each equipped with a shallow cloudlet in which APs 1, 2, and 3 are connected to the first PoP, and APs 4 and 5 to the second PoP. The network model is also assumed to have a deep cloudlet. We fix the bandwidth capacity of all the links to 1Gbps. Moreover, we consider three types of VMs (m3 large, c3 xlarge, and r3 2xlarge) and three types of resources (CPU, memory, and storage)~\cite{EC2}.
The cloudlets are assumed to be equipped with the same type of PM but different numbers of PMs are available at different hierarchical levels. The power consumption of a PM is set to 0.7kWh and the power consumption of each type of VM is estimated accordingly based on its resource demands and the resource supply of the PM. The price of electricity is fixed to 2 cent/kWh. The price of the bids are generated randomly using a triangle distribution~\cite{lampe2012maximizing} assuming that the submitted price for each type of VM will not exceed its on-demand price available at~\cite{EC2}.

\begin{figure*}[!htbp]
\center
\includegraphics[width=9cm,height=9cm]{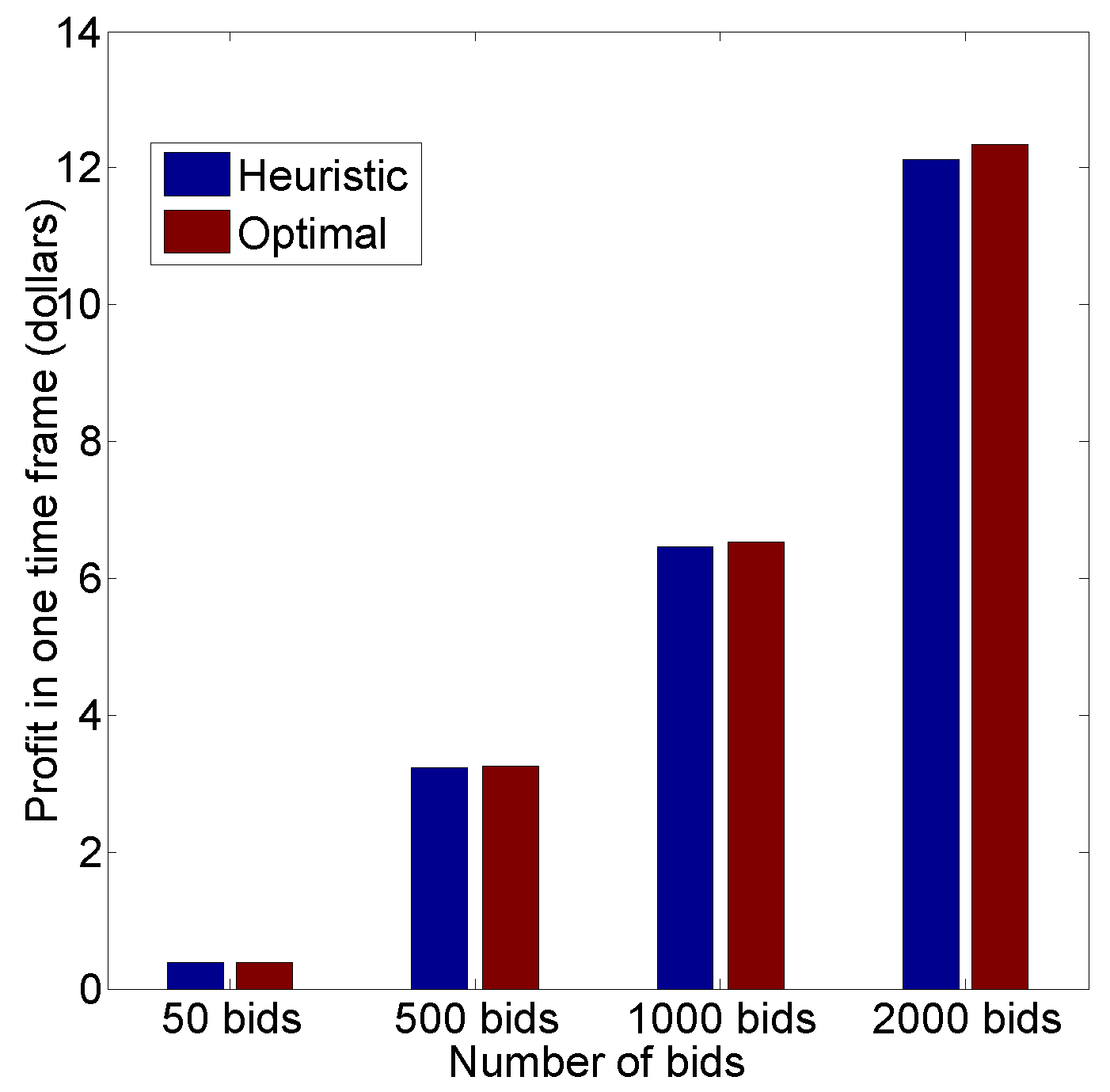}
\caption{Profit comparison between heuristic and optimal approaches for case 1.}
\label{fig:2}
\end{figure*}
\begin{figure*}[!htbp]
\center
\includegraphics[width=9cm,height=9cm]{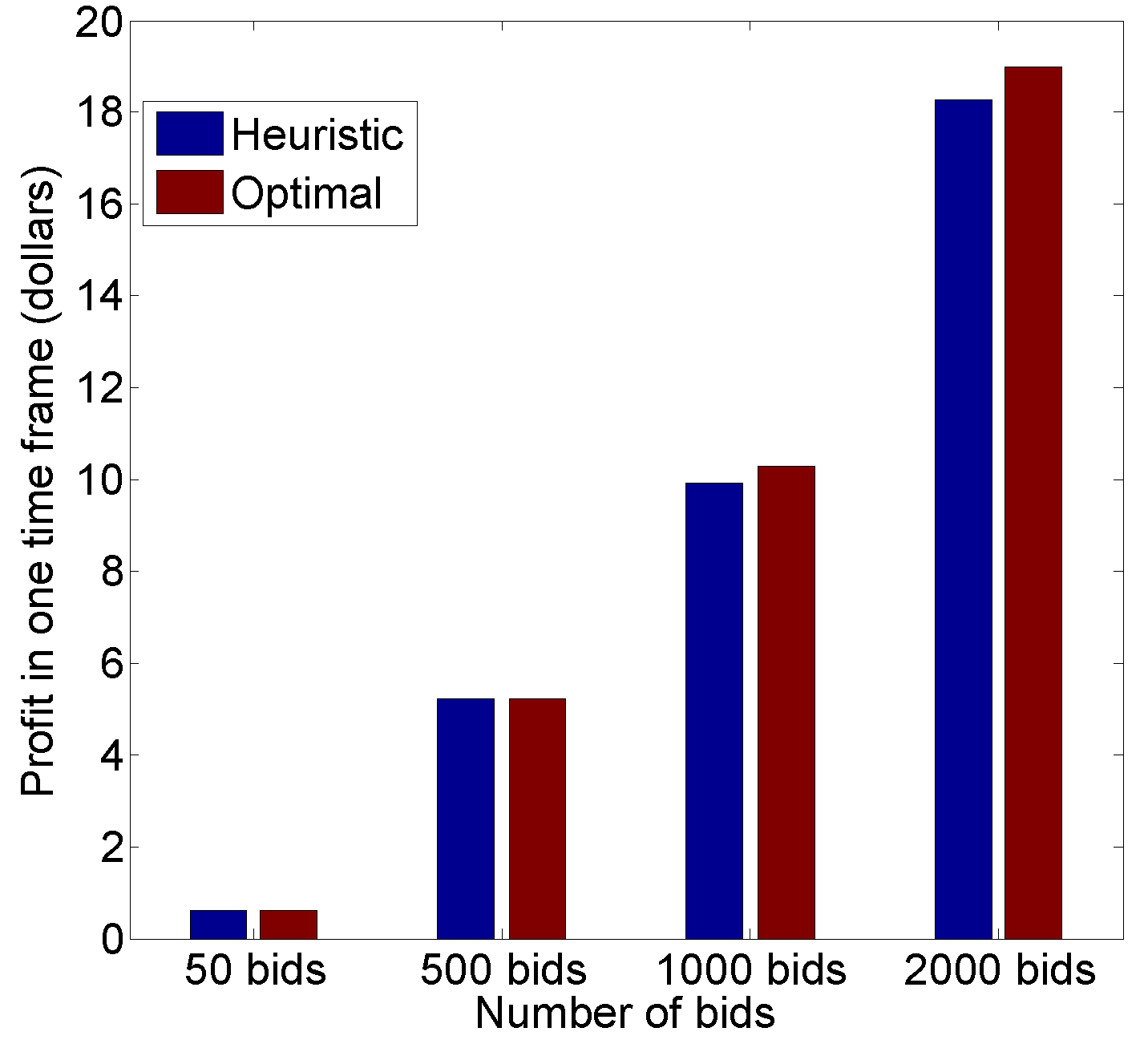}
\caption{Profit comparison between heuristic and optimal approaches for case 2.}
\label{fig:3}
\end{figure*}

\begin{figure*}[!htbp]
\center
\includegraphics[width=9cm,height=9cm]{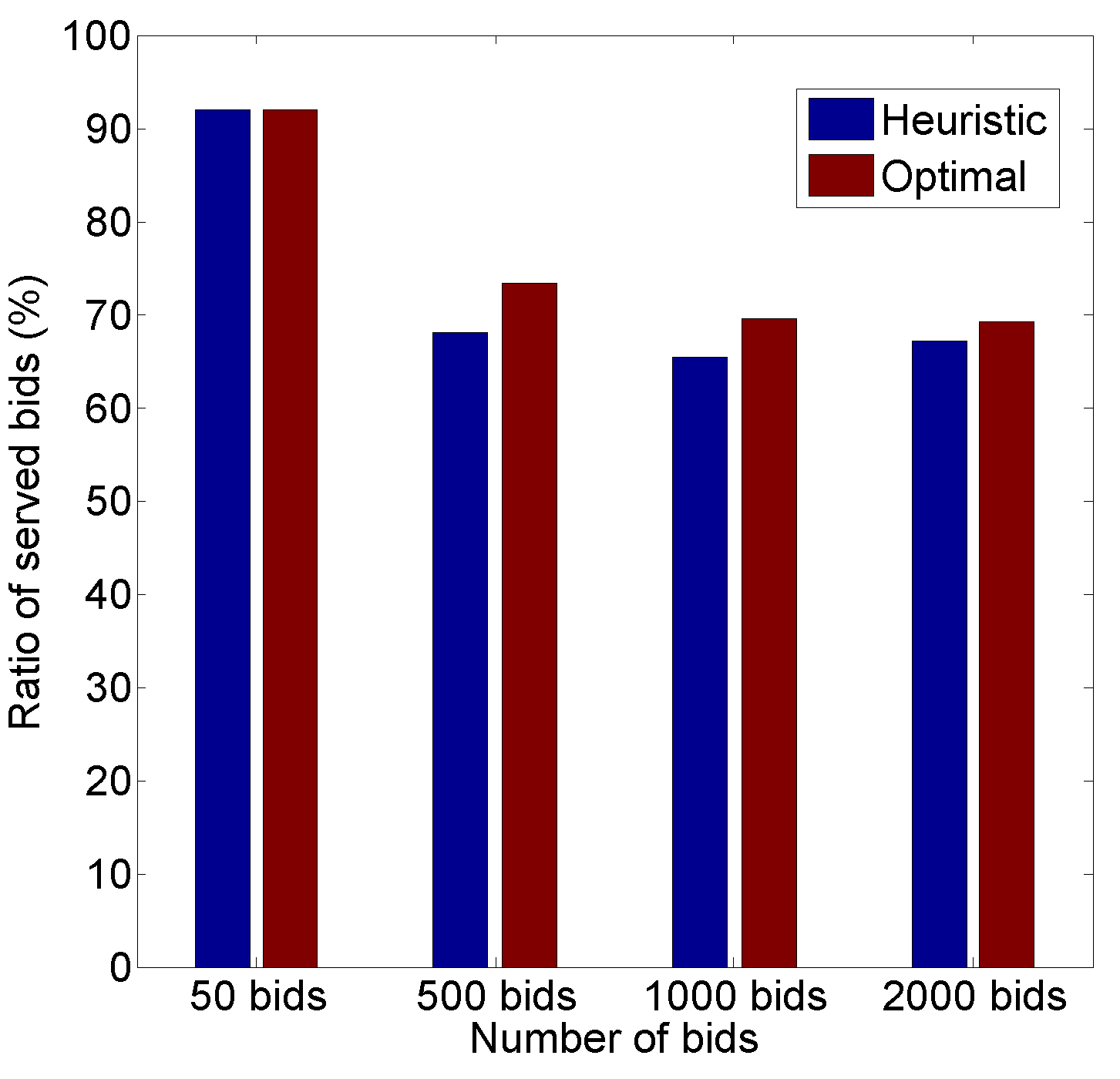}
\caption{Ratios between the served bids and the total bids for case 1.}
\label{fig:4}
\end{figure*}
\begin{figure*}[!htbp]
\center
\includegraphics[width=9cm,height=9cm]{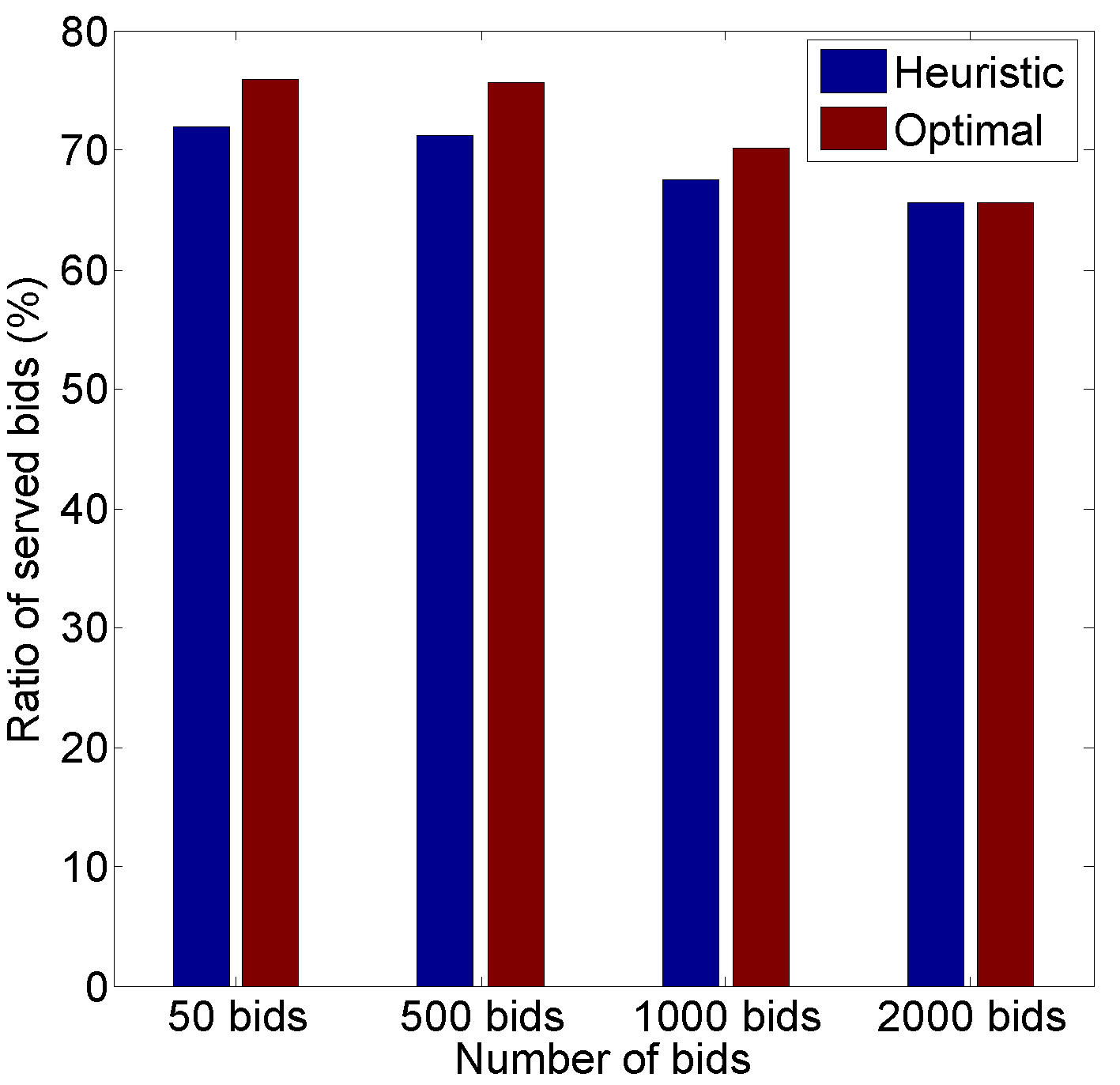}
\caption{Ratios between the served bids and the total bids for case 2.}
\label{fig:5}
\end{figure*}
\begin{figure*}[!htbp]
\center
\includegraphics[width=9cm,height=9cm]{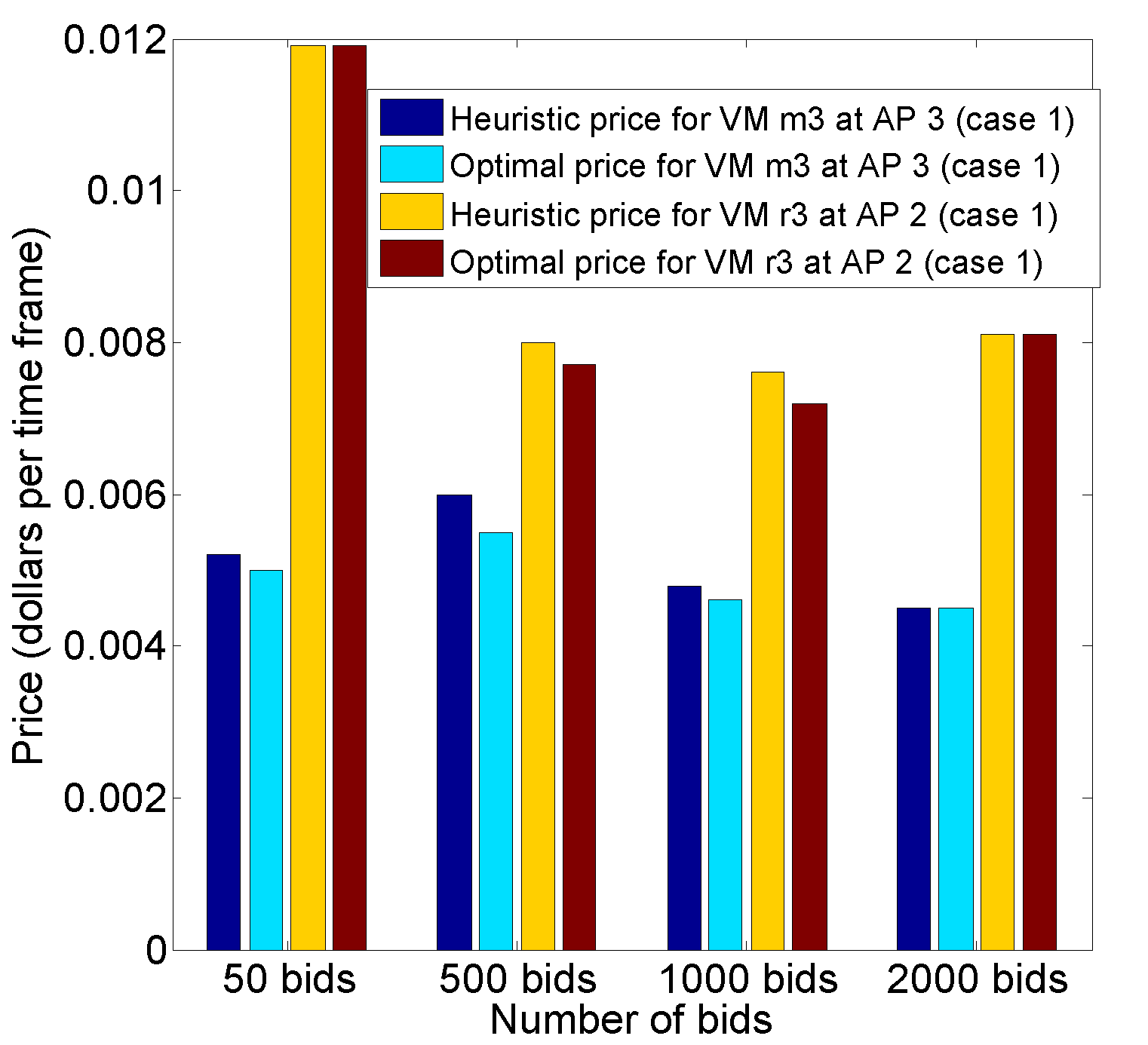}
\caption{Local prices comparison between heuristic and optimal approaches.}
\label{fig:6}
\end{figure*}

CVX~\cite{CVX} combined with Gurobi~\cite{gurobi} and MATLAB are used to simulate the BLP and the two phases heuristic approach. For performance evaluations, we study two cases, each with four different scenarios, i.e., 50, 500, 1000 and 2000 bids. In the first case study, we fix the ratio of bids submitted for three types of VMs as m3:c3:r3=2.5:1.5:1, corresponding to the case that the users are more interested in a smaller type of VM, i.e., m3. On the other hand, for the second case study, we change the ratio to m3:c3:r3=1:1.5:2:5 assuming that the users are more interested in a larger type of VM, i.e., r3. The AP locations for the bids are generated randomly in each case.

The computation time of the optimal approach (BLP) and the heuristic algorithm for different scenarios are compared in Table~\ref{table2}. While the heuristic algorithm provides the suboptimal solution within a few seconds, the computation time of the optimal approach grows fast with the number of bids. Figures~\ref{fig:2} and~\ref{fig:3} show the profits gained in one time frame for case 1 and case 2, respectively. As we can see in these figures, the heuristic algorithm results in a profit quite close to the profit of the optimal approach. We further compare the performance of the heuristic and optimal approaches by providing the ratio of the served bids in Figures~\ref{fig:4} and~\ref{fig:5} for case 1 and 2, respectively. Here, the ratio of the served bids is defined as the total number of served bids divided by the total number of submitted bids. As demonstrated in these figures, the heuristic approach serves nearly the same number of bids as the optimal approach. Finally, we validate the performance of the VM pricing algorithm in Figure~\ref{fig:6}.
Owing to similarity, we only compare two prices as examples, and we choose m3 for case 1 and r3 for case 2 since m3 and r3 are the most demanded VMs in case 1 and 2, respectively. As demonstrated in the figure, the estimated price of the heuristic VM pricing for most scenarios is slightly higher than the optimal price, i.e., the heuristic pricing serves fewer bids than the optimal one, as also confirmed by the results shown in Figures~\ref{fig:4} and~\ref{fig:5}.

\section{Conclusion}\label{sec:conclude}
In this study, we have proposed a new hierarchical architecture in the context of mobile edge computing called HI-MEC. Specifically, we have introduced the concept of field, shallow and deep cloudlets deployed in three hierarchical levels in accordance with the principle of LTE-advanced mobile backhaul network. Based on the proposed model, a two time scale optimization approach for resource allocation is introduced. In particular, a BLP is formulated to maximize an auction-based profit for concurrent VM pricing and VM distribution, and accordingly heuristic algorithms are designed to solve this problem in a reasonable time. Moreover, a convex optimization problem for bandwidth allocation is formulated and a centralized solution to this problem is derived. The proposed hierarchical model and the two time scale optimization platform have been demonstrated to effectively facilitate the resource allocation to the subscribers of a MEC network.

\bibliographystyle{IEEE}
\bibliography{ref}

\end{spacing}

\end{document}